\documentclass[twocolumn,prl,aps,superscriptaddress]{revtex4}

\usepackage{epsf}
\usepackage{graphicx}

\begin{document}

\title{Self-assembly, modularity and physical complexity}

\author{S. E. Ahnert}
\affiliation{Theory of Condensed Matter, Cavendish Laboratory, University of Cambridge, JJ Thomson Avenue, Cambridge CB3 0HE, UK}
\author{I. G. Johnston}
\affiliation{Rudolf Peierls Centre for Theoretical Physics, University of Oxford, 1 Keble Road, Oxford OX1 3NP, UK}
\author{T. M. A. Fink}
\affiliation{CNRS UMR144, INSERM U900, Institut Curie, 26 rue d'Ulm, Paris, F-75248 France}
\affiliation{Mines ParisTech, Fontainebleau, F-77300 France}
\affiliation{London Institute for Mathematical Sciences, 22 South Audley St, London W1K 2NY, UK}
\author{J. P. K. Doye}
\affiliation{Physical \& Theoretical Chemistry Laboratory, Department of Chemisty, University of Oxford, South Parks Road, Oxford OX1 3QZ, UK}
\author{A. A. Louis}
\affiliation{Rudolf Peierls Centre for Theoretical Physics, University of Oxford, 1 Keble Road, Oxford OX1 3NP, UK}

\begin{abstract}
We present a quantitative measure of physical complexity, based on the amount of information required to build a given physical structure through self-assembly. Our procedure can be adapted to any given geometry, and thus to any given type of physical system. We illustrate our approach using self-assembling polyominoes, and demonstrate the breadth of its potential applications by quantifying the physical complexity of molecules and protein complexes. This measure is particularly well suited for the detection of symmetry and modularity in the underlying structure, and allows for a quantitative definition of structural modularity. Furthermore we use our approach to show that symmetric and modular structures are favoured in biological self-assembly, for example of protein complexes. Lastly, we also introduce the notions of joint, mutual and conditional complexity, which provide a useful distance measure between physical structures.
\end{abstract}

\maketitle

\section{Algorithmic complexity}
More than forty years ago, Kolmogorov \cite{Kolmogorov} and Chaitin \cite{Chaitin} laid the foundations of algorithmic information theory, by introducing the concept of algorithmic information content, or Kolmogorov complexity, for a given string of information \cite{CoverThomas}. This measure of complexity is defined as the length of the shortest possible program on a universal computer that will output the string in question. 
Here we propose a conceptually analogous measure of the complexity of any connected physical structure. Instead of a universal computer which translates a program into a string of information, we consider a general framework of self-assembly rules, which act together to create a physical object. The `program' now is our set of self-assembly building blocks and rules, the `computer' is given by the physical interactions of the self-assembling building blocks, and the `output' is the final structure. Using this approach we investigate the physical complexity of shapes in two and three dimensions, including polyominoes, molecules and protein complexes. Our work generalizes ideas first explored in \cite{Winfree00,Winfree06}, and opens them up to a wide range of applications. Furthermore, in the context of protein complexes it offers the kind of biological application of information-theoretic concepts demanded in \cite{Adami}.

\section{Self-assembly kit}
There are many examples of self-assembling structures in physics, chemistry and biology \cite{Whitesides}. Examples include thin films \cite{Krausch}, micelles \cite{Israelachvili}, viruses \cite{Fraenkel-Conrat,Zlotnick} and DNA \cite{Winfree98,Mao,Jaeger,Goodman,Rothemund,Winfree08}. Our aim is to introduce a general framework for the theoretical study of self-assembling structures. This framework can be used to study the properties of real self-assembling systems, but, more generally, it can also be used to measure the physical complexity of any construct, self-assembling or not. The exact nature of the self-assembly framework depends on the underlying physical system, but it always contains two basic ingredients: a set of building blocks and a set of rules. We shall call this combination an {\em assembly kit} $S$. Each building block $i$ has $f_i$ interfaces, which typically are subject to geometric constraints (depending on the physical system). Attached to each interface $j$ of a given building block $i$ is an integer $\chi_{ij} \in [1, ...,c]$. The $c$ possible values of these integers are the {\em colours} of these interfaces. The number of distinct colourings of the building blocks depends entirely on the geometry of the problem. The second ingredient of the assembly kit is the set of rules, which takes the form of an interaction matrix between colours. In the simplest case this matrix is binary, where 1 signifies attraction and 0 signifies no interaction at all. Many more sophisticated interaction matrices involving repulsion and a continuous spectrum of energies are easily imaginable.  

\begin{figure}[t!]
\scalebox{0.25}[0.25]{\rotatebox{0}{\epsfbox{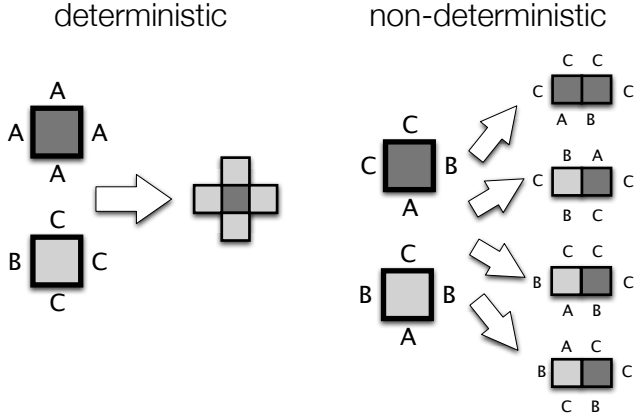}}}
\caption{An example of deterministic and non-deterministic self-assembly kits, using simple 2D lattice structures (polyominoes). In both cases, colours A and B attract each other, but C attracts neither A nor B. No colour attracts itself. The kit on the left will always assemble into the cross shape while that on the right will assemble into an irregular cluster, as there are several ways in which the two blocks can attach.}\label{crossblob}
\end{figure}

For any system of self-assembling particles we need to also specify a model for the actual assembly process. A convenient choice is a model assuming a single nucleus in solution \cite{Winfree00}, which makes the assumption that each disjoint object has one fixed nucleus building block which is surrounded by a solution containing a freely moving population containing many copies of each type of building block. Each time step (i) a fixed building block, (ii) a site adjacent site to it, (iii) a random rotational orientation, and (iv) a building block from the solution are chosen at random, and the new, randomly rotated building block becomes fixed to its position if the rules allow it. Note that some assembly kits always assemble into the same shape - these we call `deterministic' - while ones which contain ambiguous rules are `non-deterministic'. See Figure \ref{crossblob} for an example of a deterministic and a non-deterministic self-assembly kit.

As a simple example of a self-assembling system, we will consider self-assembling {\em polyominoes}. A polyomino (also known as a {\em lattice animal}) is a set of connected sites on a (typically square) lattice \cite{polyominoes}. These connected sites are our self-assembly building blocks. Every building block has four sides (so that $f_i = 4$ for all $i$), which are painted with one of $c$ colours. These colours can attract each other or not, as encoded in a $c \times c$ binary interaction matrix. Each distinct way of colouring a building block corresponds to a different building block {\em type}. We do not regard rotated colourings as distinct. The geometry of the 2D lattice gives rise to a particular set of building block colourings in the context of self-assembly. If we have $c$ colours, the total number of such colourings is \cite{necklaces}:
\[
N_c = (c^4 + c^2 + 2c)/4
\]
These particular colourings are also known as {\em necklaces}, which can be defined as equivalence classes of strings under rotation \cite{necklaces}. The definition of necklaces used here assumes that the building blocks have a fixed chirality - in other words that the necklaces which the colours form on the building blocks are {\em fixed}.\footnote{For {\em free} necklaces, which represent building blocks with no fixed chirality there are $M_c = (c^4 + 2c^3 + 3 c^2 + 2c)/8$ necklaces \cite{necklaces}. In general we will assume fixed chirality.} 

\section{the minimum kit}

Every deterministic assembly kit $S_A$, which always assembles into a structure $A$, requires a certain amount of information $I(S_A)$ to describe it in some given language $L$. Our aim is to minimize this quantity, as we define the length of the description of the {\em minimum assembly kit} $\tilde{S}_A$ as the {\em complexity} $K(A)$ of structure $A$:
\[
K(A) = I(\tilde{S}_A) = \min_{S_A} I(S_A)
\]
in analogy to the concept of Kolmogorov complexity. Any symmetry or modularity which the structure $A$ contains decreases the amount of information required to describe the structure and will therefore be reflected in its minimum assembly kit $\tilde{S}_A$, and by extension in the value of $K(A)$.

If a minimum assembly kit is deterministic, an interaction matrix $A$ (with elements $a_{ij}$) between a total of $c$ colours, of which $c_s$ self-interact, can be rewritten as:
\[
a_{ij} = [1 - (i \, {\rm mod} \, 2)] \delta_{i(j+1)} + (i \, {\rm mod} \, 2) \delta_{i(j-1)} 
\]
for $i \le c-c_s$, and $a_{ij} = \delta_{ij}$ otherwise, so that one colour always only interacts with one other colour. With this constraint, the amount of information, in bits, required to describe a self-assembly kit $S_A$, with $b$ building block types, is:
\begin{equation}\label{info}
I(S_A) = \log_2 (c_s + 1) + \sum_{i = 1}^b c_i \log_2 c + \log_2 F_i
\end{equation}
The first term relates to the number of self-interacting colours, the second measures the information required to describe which $c_i$ colours out of the total of $c$ colours appear on building block $i$, and the third term $\log_2 F_i$ measures the information describing the distinct arrangement of the $c_i$ colours on the $f_i$ faces of building block $i$. For a general building block with $f_i$ labelled faces, $F_i$ takes the form of:
\[
F(c_i, f_i) = \sum_{k_1=1}^{f_i-c_i+1} \sum_{k_2=1}^{f_i-c_i+2-k_1} \, ... \, \sum_{k_{c_i-1}=1}^{f_i-c_i+(c_i-1)-\Sigma'} {f_i ! \over \prod_{m=1}^{c_i} k_m!}
\]
where $\Sigma' = \sum_{j=1}^{c_i-2} k^{(i)}_j$, and the $k^{(i)}_j$ signify the number of times colour $j$ occurs on block $i$. 

For polyominoes $F_i = F(c_i) = N'_{c_i}$, where $N'_{c_i}$ is the number of necklaces with {\em exactly} $c_i$ colours, given by 
\[
N'_{c_i} = N_{c_i} - \sum_{k = 1}^{c_i-1} \left(\begin{array}{c}c_i\cr k\end{array}\right)N'_k
\]
with $N'_1 = 1$. It follows that $N'_2 = 4$, $N'_3 = 9$, and $N'_4 = 6$. As before, the complexity $K(A)$ of polyomino $A$ is the minimum of $I(S_A)$ over all possible assembly kits $S_A$. 
Note that Wang tiles \cite{Wang} are a special case of self-assembling polyominoes. The tile system described in \cite{Winfree06} is also similar to our framework for the case of polyominos, but (like Wang tiles) only considers self-interacting colours, and treats rotated tiles as distinct. As a result our encoding, based on necklaces, makes symmetry and modularity in the structure more directly measurable. 


If the faces are geometrically unconstrained - as one would imagine for a node with a set of freely moving links - and hence unlabelled, we would only need to specify how much there is of each colour. This can be written using $F_i = \prod_j^{c_i} k^{(i)}_j$, so that $\log_2 F_i = \sum_j^{c_i} \log_2 k^{(i)}_j$. However, this only works under the condition that multiple connections between the same pair of building blocks are prohibited.




The general algorithm we use to find the minimum assembly kit $\tilde{S}$, and thus the complexity $K$, for polyominoes and other structures is described in the following section. 

\section{A general measure of structural complexity}


Below we describe a general algorithm for minimizing the assembly kit size for a connected physical structure without relying on steric effects. Taking these into account can minimize the assembly kit even further, but their computation is highly dependent on the geometry of the system and in most cases non-trivial (see Discussion). Note also that in some structures, such as polyominoes, some edges of the contact graph can be redundant in the context of the assembly process. Whether contact graph edges in general can be redundant or not depends on the nature of the structure and the assumptions connected to the self-assembly of that structure (see Discussion). Similarly, when interfaces are defined by geometry, as for the four sides of a polyomino building block, it makes sense to introduce a neutral colour ($\nu = 1$ below). In systems with a varying number of interfaces on the building blocks, neutral colours are usually not required ($\nu = 0$).

\noindent
To minimize the assembly kit we take the following steps:
\begin{enumerate}
\item{Divide the structure into building blocks (usually a natural division). The number of building blocks is the {\em size} of the structure, denoted $z$.}
\item{Determine the equivalence of these units in terms of any additional criteria (e.g. types of atoms, proteins). This categorization is the $species$ of building block.}
\item{Establish a contact graph $a_{ij}$ for the units (in some cases, such as molecules, this may require setting a distance cutoff).}
\item{{\em If edges can be redundant:} Consider the space of all spanning subgraphs of this graph.} 
\item{For the contact graph (in the case of no redundant edges) or each subgraph (if redundant edges exist):}
\begin{enumerate}
\item{Classify the (sub)graph according to the number of connections and (depending on the geometry) the arrangement of connections.}
\item{Label all nodes which are not yet labelled and which have exactly one unlabelled node among their neighbours. The new labels distinguish nodes according to their species as well as the topologically distinct label distributions among their neighbours.}\label{rep}
\item{Repeat step \ref{rep} until all nodes are labelled or no more nodes can be labelled.}
\item{All labelled nodes we define as {\em category 1} nodes and any remaining unlabelled nodes (i.e. nodes with at least two unlabelled neighbours) are defined as {\em category 2} nodes.}
\item{Label all category 2 nodes simultaneously according to their neighbourhoods.}\label{rep2}
\item{Repeat step \ref{rep2}, using the previous labellings to distinguish neighbourhoods, until labellings are stable.}
\item{These final labels, for nodes in both categories, denote the building block {\em types}. The number of final labels, or types, is $b$. These can be subdivided in to $b_1$ category 1 building block types and $b_2$ category 2 building block types. The category 2 type of block $i$ is denoted $t_i$.}
\item{The degree of each building block type $i$ in the contact graph (or subgraph) is the number of its interfaces $f_i$.}
\item{The total number of colours, including $\nu \in \{0,1\}$ neutral colours, is $c = 2(b_1-1)+\nu+\sum_{i,j=1}^{b_2} \left(1-\prod_{k,l=1}^{z} (1-(a_{kl} \delta_{it_k} \delta_{jt_l}))\right)$. The sum expression gives the number of different types of interfaces which occur between category 2 building block types\footnote{Heterogeneous interfaces are double-counted as, unlike homogeneous interfaces, they require two colours.}. The number of colours $c_i$ on building block $i$ is equal to the number of building block types in its contact graph neighbour set.}
\item{Using $b$, $c$, $\{f_i\}$ and $\{c_i\}$ in equation (\ref{info}), calculate the information $I$ required to specify this assembly kit, and thus the complexity $K$ of the structure.}\label{finish}
\end{enumerate}
\item{{\em If edges can be redundant:} Minimize this quantity over all spanning subgraphs.}
\end{enumerate}

\begin{figure}[t!]
\hspace{-0.3cm}\scalebox{0.24}[0.24]{\rotatebox{0}{\epsfbox{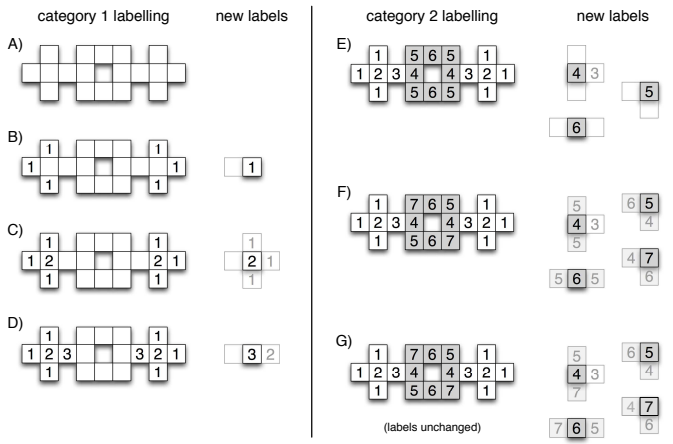}}}
\caption{An illustration of the crucial steps 5b to 5j of the algorithm for minimizing the assembly kit size, in this case for a polyomino. In every iteration of category 1 labellings (LEFT), all unlabelled nodes with exactly one unlabelled neighbour are given labels which distinguish them according to their topologically distinct neighbourhoods of unlabelled and labelled tiles. This procedure is repeated until no more blocks can be labelled in this way. The remaining blocks are given category 2 labellings (RIGHT) which are applied simultaneously, with each label distinguishing the topological neighbourhoods of the tiles in the previous iteration. Note that in the last iteration the labellings have stabilized, and only the interfaces of the building block types are updated. For structures in which edges can be redundant, this operation can be performed for all spanning subgraphs of the structure's connectivity graph, which further reduces the complexity. (In polyominoes, edges can be redundant, but there are no spanning subgraphs in the above example.)}\label{algo}
\end{figure}

\begin{figure}[t!]
\scalebox{0.5}[0.5]{\rotatebox{0}{\epsfbox{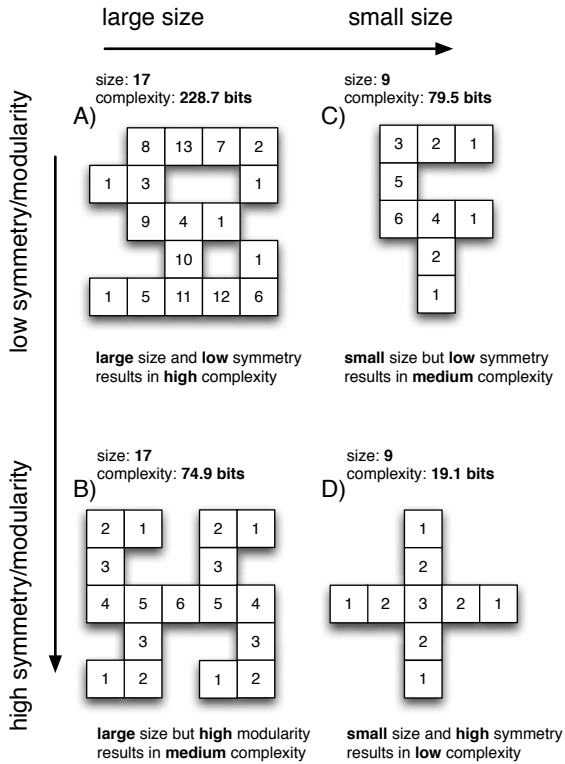}}}
\caption{The complexity values of these four polyomino shapes illustrate why the self-assembly approach is an effective way of measuring symmetry and modularity without requiring prior assumptions. If two shapes are of equal size, the one with more symmetry and modularity has a lower complexity value - compare A with B, and C with D. If on the other hand, two shapes are of similar {\em complexity}, but of different size, the larger one will be more symmetric or modular (compare B and C).}\label{minimumkit}
\end{figure}

Figure \ref{algo} illustrates the crucial steps \ref{rep} to \ref{finish} for a polyomino. Figure \ref{minimumkit} illustrates how the complexity value $K$ reflects symmetry and modularity present in the structure.

\section{Applications}

The self-assembly approach can be used to calculate complexity values for any physical structure. In order to demonstrate the broad range of potential applications we determine the complexity of (a) molecules and (b) protein complexes. 

\begin{figure}[t!]
\scalebox{0.7}[0.7]{\rotatebox{0}{\epsfbox{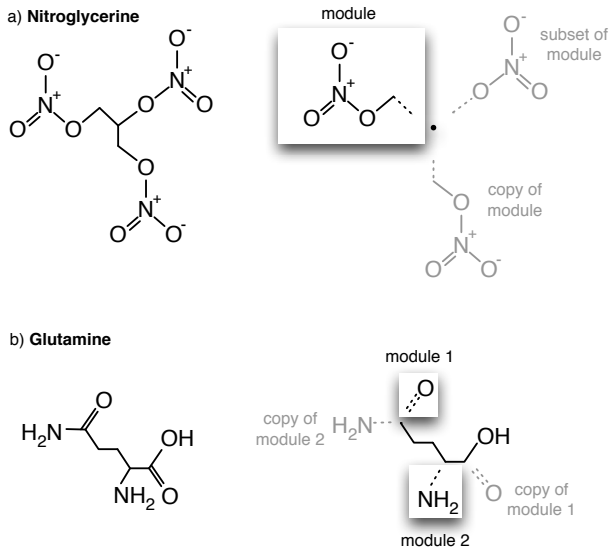}}}
\caption{Measuring the complexity of molecules -- The explosive nitroglycerine (top) and the amino acid glutamine (bottom) both consist of 20 atoms, but differ greatly in complexity. The highly modular structure of nitroglycerine with its three $\rm NO_3$ groups means that its complexity value $K$, at 52.2 bits, is little more than half that of glutamine ($K = 91.0$ bits). Note that nitroglycerine does not have simple three-fold symmetry, but a more subtle modular structure, which the self-assembly approach fully reveals. Note that we do not consider neutral colours in this structure ($\nu = 0$).}\label{nitro}
\end{figure}

The problem of molecular complexity has been studied extensively over the past seventy years, starting with work by P\'olya \cite{Polya} and Rashevsky among others \cite{Rashevsky,Trucco}, and culminating in a seminal paper by Bertz \cite{Bertz}. These approaches are based on Shannon entropy rather than algorithmic information theory and focus on symmetries rather than the more general concept of modularity. 
In molecules, we take atoms to be the building blocks and chemical bonds to be their interfaces. Simple molecules, such as those in Figure \ref{nitro}, for which we are only interested in the bond connectivity, are an example of a structure in which none of the edges can be regarded as redundant. This is because, unlike for polyominoes, we are not assuming any inherent geometry for the building blocks. If two atoms play the same self-assembly role but represent atoms of different atomic species, they must be differentiated. This also goes for atoms connected by different bond types. For example, in glutamine (see Figure \ref{nitro}), the oxygen atom connected with a double bond is a leaf of the self-assembly tree just like any of the (implicit) hydrogen atoms, but it requires a separate building block. The two molecules in our example of Figure \ref{nitro} are the amino acid glutamine and the explosive nitroglycerine, which both consist of 20 atoms. Nitroglycerine however exhibits a much higher degree of modularity, with its three $\rm NO_3$ groups, and therefore has a much lower complexity of $K = 55.3$ bits than the glutamine, for which the value is $K = 94.7$ bits. Note that nitroglycerine does not exhibit simple three-fold symmetry, but a more subtle, hierarchical modularity. Such structural features would be harder to discover using traditional approaches to the measurement of molecular complexity \cite{Rashevsky,Trucco,Bertz}, which do not take a self-assembly perspective and rely on Shannon entropy rather than Kolmogorov complexity as a measure of complexity.

\begin{figure}[t!]
\scalebox{0.65}[0.65]{\rotatebox{0}{\epsfbox{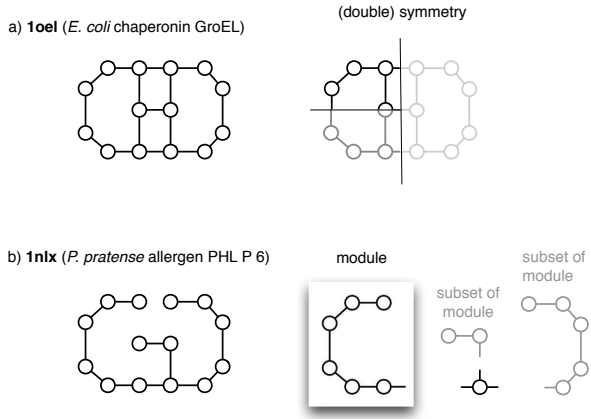}}}
\caption{We measure the complexity of two protein complexes, with PDB identifiers 1oel (a chaperonin, top) and 1nlx (an allergen, bottom), which have 14 proteins each. The symmetry of the chaperonin complex means that it has a much lower complexity value of $K = 31.5$ bits, compared to $K = 50.2$ bits for the allergen complex. Note that we are assuming non-redundant edges in this calculation, so that all building blocks of the chaperonin complex are category 2 and all building blocks of the allergen complex are category 1. Furthermore we do not consider neutral colours ($\nu = 0$), and in the case of the chaperonin complex we have three self-interacting colours ($c_s = 3$). Note also that both complexes are homomers, i.e. they only have one type of subunit.}\label{proteincomplex}
\end{figure}

Many important biochemical structures are \emph{protein complexes}, consisting of several individually formed and folded protein \emph{subunits} bound together to produce functional cellular machinery. These subunits may include different types of protein and several copies of the same protein. The physical structure of protein complexes, as with protein themselves, is important in determining the functionality of the complex. The manner in which the subunits bond to form the final complex is known as the \emph{quaternary structure} of the complex. The 3DComplex database\cite{3DComplex} contains a description of the quaternary structures of thousands of protein complexes, in terms of subunit type and inter-subunit bonding. 
If we have two proteins which play the same role in the self-assembling structure but are different proteins, we can choose to count them as two different building blocks (analogous to the aforementioned distinction between atomic species in molecules). In the following analyses we are only interested in the connectivity of proteins (equivalent to the QS Topology level in the 3DComplex database), and therefore do not distinguish between different proteins. The two protein complexes in our example of Figure \ref{proteincomplex} are a chaperonin complex ({\em E. coli} chaperonin GroEL; PDB identifier: 1oel) and an allergen complex ({\em P. pratense} allergen PHL P 6; PDB identifier: 1nlx). Both consist of 14 proteins, but the former displays a much higher degree of symmetry and a much lower complexity value of $K = 31.5$ bits, versus $K = 50.2$ bits for the allergen (which is still somewhat modular). 

\begin{figure}
\begin{center}
\includegraphics[width=8cm]{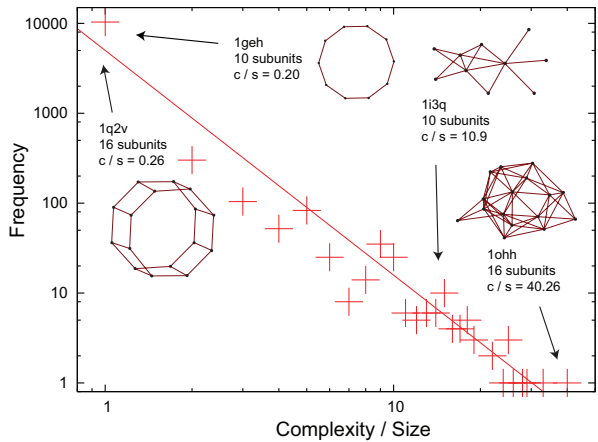}
\caption{(Colour online) Histogram of protein quaternary structure assembly complexity with frequency of occurrence in the 3DComplex database. Insets illustrate two pairs of equally sized structures with high and low complexity values. 1geh, 1i3q, 1q2v, and 1ohh are the PDB identifiers of the complexes. The plot has an $R^2 = 0.93$ correlation with a power law decay. Note that in this case we do not distinguish between different types of subunit.}\label{iaincomplex}
\end{center}
\end{figure}

More complex protein structures require more unique inter-subunit bonds types, compared to less complex structures which can re-use bonds and be constructed through simple repetition of subunits. As an increase in bond types corresponds biologically to the presence of more unique bonding sites on subunit proteins, more complex protein structures can be thought of as requiring more evolutionary innovation to produce and would therefore be expected to occur less frequently in biological organisms \cite{Villar,Levy}. This hypothesis is confirmed by Figure \ref{iaincomplex}, which shows a histogram of complexity values -- normalized by the size of the protein complex, to avoid size effects -- for the 15733 protein complexes in the 3DComplex database \cite{3DComplex}. The distribution closely ($R^2= 0.93$) follows a power-law decay. 

In both of these cases - molecules and protein complexes - we assume geometrically unconstrained faces for the building blocks; in other words, we use $F_i = \prod_j^{c_i} k_j$. While the chemical bonds of atoms and the interfaces of proteins are in fact usually constrained, this information is not part of the structural formula of the molecule or the contact graph of the protein complex. If this additional level of resolution is required, a more realistic self-assembly model can be constructed, based on the exact three-dimensional characteristics of the atoms or proteins, and using the $F(c_i, f_i)$ term specified above.

\section{Modularity}

The self-assembly perspective provides an intuitive definition of the modularity of a structure: If part of the structure appears several times, it still only needs to be encoded once. This is why modularity and symmetry (being a special case of modularity) lead to more efficient self-assembly kits and a lower value of the complexity measure $K$. Formally we can define the modularity $m$ of a structure of size $z$ as the average number of times one of the $b$ different building block types in the minimum assembly kit is used in the structure, which is simply:
\[
m = {z \over b}
\]
We can furthermore define a {\em module} formally as a connected set of building blocks which appears more than once in a given structure. Note that modules can overlap: A subset of a module could form another module, appearing a different number of times than the whole module. The molecule in Figure \ref{nitro}a illustrates such a case. 

\begin{figure}
\begin{center}
\includegraphics[width=8cm]{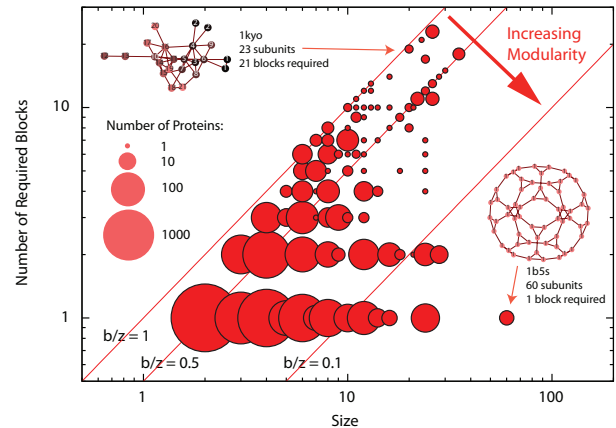}
\caption{(Colour online) The position of the 15733 protein complexes from \cite{3DComplex} in the space of $b$ (number of building block types) and $z$ (size of the complex). Many protein complexes are highly modular, and this is true across a wide range of sizes. In this plot complexes of equal modularity $m = z/b$ lie on a diagonal line with positive gradient. The lines are shown for $m = $1, 2, and 10 ($b/z = $1, 0.5, and 0.1). The sizes of the circles show how many complexes lie at a given position $(z,b)$. The insets show two examples (with PDB identifiers 1kyo and 1b5s), with high and low modularities.}\label{modular}
\end{center}
\end{figure}

The majority of protein complexes in the 3DComplex database show high modularity values (Figure \ref{modular}) with a common trend observable along the $b/z = 0.5$ line, indicating many proteins consist of structures involving two copies of all constituent subunits.

To further illustrate how the complexity $K$ and the modularity $m$ measure the physical complexity of protein complexes, we consider two of the outliers in the complexity and modularity histograms, the high-complexity 1ohh (Figure \ref{iaincomplex}) and high-modularity 1b5s (Figure \ref{modular}). 1ohh consists of two copies of bovine F$_1$-ATPase (itself a protein complex) in complex with its regulatory protein IF$_1$\cite{cabezon2003structure}. The regulatory protein binds simultaneously to both copies of the main complex, but slightly asymmetrically, leading to asymmetric interactions being recorded in the 3DComplex database. This asymmetry results in extra information being required to describe the combined quaternary structure, and the observed high complexity value.
1b5s is a multienzyme complex consisting of multiple copies of dihydrolipoyl acetyletransferase (E2p)\cite{izard1999principles}. The E2p protein has the potential to occupy quasi-equivalent positions, as seen in virus structures\cite{caspar1962principles}, and is also observed to form cubic complexes. The highly-modular, dodecahedral structure exhibited in 1b5s is an efficient way of grouping many copies of an active protein in a geometry that facilitates enzymatic activity: the large windows in the structure allow passage of the substrate and product between the inner cavity and the substrate. The structure of the protein subunits allows this structure to be realised with just one building block type, resulting in high modularity.

\section{Joint, Conditional, and Mutual Complexity}

If we have two structures $A$ and $B$ with minimum assembly kits $\tilde{S}_A$ and $\tilde{S}_B$, then the {\em joint} minimum assembly kit $\tilde{S}_{A,B}$ is the minimum kit which can assemble both structures if an appropriate subset of building blocks is chosen. The amount of information required to describe this kit is the {\em joint complexity} $K(A,B)$ of $A$ and $B$. This definition can easily be generalized to more than two structures. 

Let us define $\tilde{S}'_A$ as the subset of $\tilde{S}_{A,B}$ which forms structure $A$, and $\tilde{S}'_B$ as the subset of $\tilde{S}_{A,B}$ which forms structure $B$ (note that e.g. $\tilde{S}_A$ is not necessarily equal to $\tilde{S}'_A$ due to the colour minimization), so that $\tilde{S}_{A,B} = \tilde{S}'_A \cup \tilde{S}'_B$. Furthermore, let us define the {\em conditional} minimum assembly kit $\tilde{S}_{A|B}$ as the set of building blocks we need in addition to $\tilde{S}'_B$ in order to form structure $A$. Then we can write: 
\[
\tilde{S}_{A|B} = \tilde{S}_{A,B} \backslash \tilde{S}'_{B}
\]
where $\backslash$ denotes the set theoretic difference operation. The definition of $\tilde{S}_{B|A}$ follows accordingly. Hence we can also define a {\em conditional complexity} $K(A|B)$, which is the amount of information needed to describe the building blocks in $\tilde{S}_{A|B}$. Because the way we describe the assembly kit is additive in the number of building blocks, we can write 
\[
K(A|B) = K(A,B) - K'(B)
\]
since $K'(B)$ is the information required to describe the building blocks in $\tilde{S}'_B$. The relationship between $K(B)$ and $K'(B)$ is given by
\[
K'(B) = K(B) + \sum_i c_i \log_2 {c_{A,B} \over c_B}
\]
where $c_{A,B}$ is the total number of colours in $\tilde{S}_{A,B}$ and $c_B$ is the total number of colours in $\tilde{S}_B$. Because of the minimization of colours, $c_{A,B} = \max(c_A, c_B)$. Hence, if $c_B \ge c_A$, then $K'(B) = K(B)$.

Similarly, we can define a mutual minimum assembly kit $\tilde{S}_{A:B}$, which corresponds to the intersection 
\[
\tilde{S}_{A:B} = \tilde{S}'_A \cap \tilde{S}'_B = \tilde{S}'_A \backslash \tilde{S}_{A|B} = \tilde{S}'_B \backslash \tilde{S}_{B|A}
\]
From this follows the {\em mutual complexity}
\begin{eqnarray}\nonumber
K(A:B) &=& K'(A) - K(A|B) = K'(B) - K(B|A)\cr &=& K'(A) + K'(B) - K(A,B)
\end{eqnarray}

\begin{figure}[t!]
\hspace{-0.5cm}\scalebox{0.45}[0.45]{\rotatebox{0}{\epsfbox{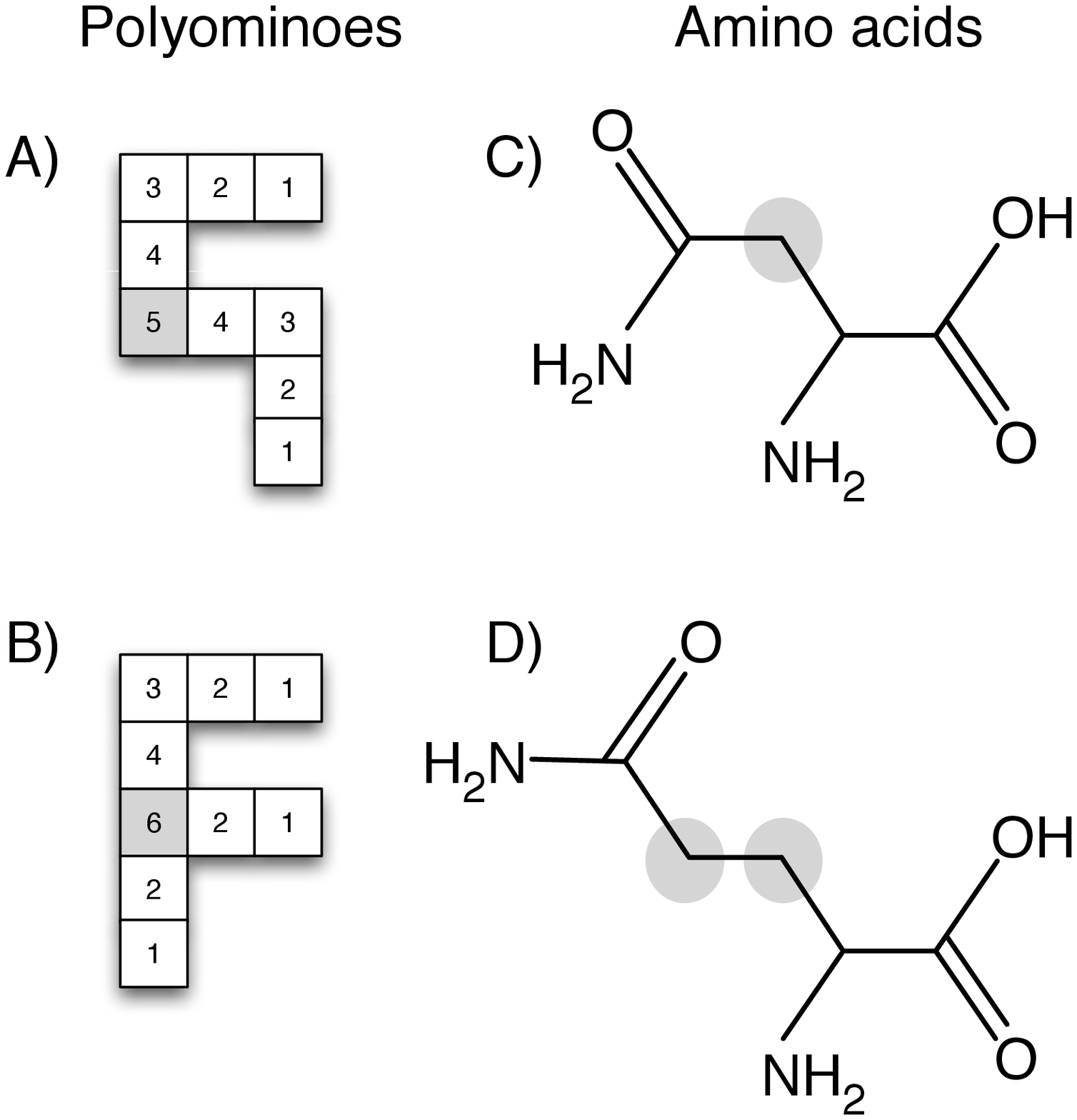}}}
\caption{POLYOMINOES (left): The two polyominoes share many building block types, with the only two unique ones being blocks 5 and 6 (marked in grey). Hence, the joint set is $\tilde{S}_{A,B} = \{1,2,3,4,5,6\}$, the mutual set is $\tilde{S}_{A:B} = \{1,2,3,4\}$ and the conditional sets are: $\tilde{S}_{A|B} = \{5\}$ and $\tilde{S}_{B|A} = \{6\}$. Building block 5 contributes $K(A|B) = 2 \log_2 9 + 2 = 8.4$ bits to the complexity $K'(A)$ of the A shape, while block 6 contributes $K(B|A) = 4 \log_2 9 = 12.7$ bits to $K'(B)$. It follows therefore that the joint complexity is $K(A,B) = 67.4$ bits and the mutual complexity is $K(A:B) = 46.4$ bits, compared to the standalone values of $K(A) = K'(A) = 54.7$ bits and $K(B) = K'(B) = 59.1$ bits (see Figure \ref{minimumkit}). AMINO ACIDS (right): The two amino acid molecules asparagine (top, C) and glutamine (bottom, D) share the amino ($\rm NH_2$) and carboxyl ($\rm CO_2H$) groups common to all amino acids, as well as the carboxamide group ($\rm CONH_2$). In a self-assembly framework these two structures have complexities of $K(Asn) = 74.3$ bits and $K(Gln) = 91$ bits. While $K'(Gln) = K(Gln)$, we have $K'(Asn) = 78.0$ bits. Because the two molecules share three groups, their joint complexity is not much larger than their individual complexities, at $K(Asn,Gln) = 104.0$ bits, and their mutual complexity is not much smaller, at $K(Asn:Gln) = 65$ bits, than the complexities of the individual molecules. Their conditional complexities are correspondingly low, at $K(Asn|Gln) = 13$ bits and $K(Gln|Asn) = 26$ bits. The conditional complexities give the amount of information required to describe the building blocks (atoms) which are unique (in their self-assembly role) to the given amino acid. These atoms are marked with grey circles.}\label{joint_examples}
\end{figure}

In order to account for the relative sizes of the structures we compare using these measures, we can define relative versions of the above quantities. These are {\em relative conditional complexity}:
\[
K^{rel}(A|B) = {K(A|B) \over K'(B)}
\]
and the {\em relative mutual complexity} 
\[
K^{rel}(A:B) = {K(A:B) \over K(A,B)}
\]
Note that the latter measure resembles the Jaccard index \cite{jaccard01}. For an illustration of joint, mutual and conditional complexity, see Figure \ref{joint_examples}.

\section{Discussion}

\begin{figure}[t!]
\scalebox{0.45}[0.45]{\rotatebox{0}{\epsfbox{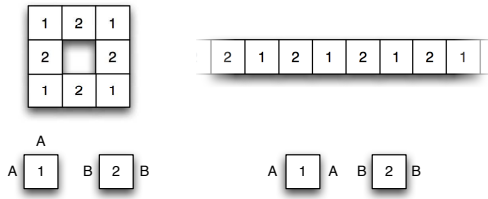}}}
\caption{A simple example of a steric effect. The two blocks 1 and 2 have colours A and B on their interfaces. These colours attract each other. All other faces are neutral. Certain arrangements of colours will lead to self-delimiting structures purely because of the geometry of the building blocks. The complexity of such structures can be taken to be the same as that of an infinite chain consisting of the same sequence of blocks, but only if each loop structure inside a bigger structure has a distinct (set of) species of building blocks.}\label{steric}
\end{figure}

{\em Steric effects} -- For structures which contain loop structures formed by repeating units, it is possible to exploit steric effects in order to reduce the size of the assembly kit below the minimum size found by our algorithm (which explicitly excluded such effects in its definition). An example of a steric effect would be a polyomino which is self-limiting in a deterministic way, purely because of the geometric constraints of the building blocks. As long as each distinct type of loop structure is formed by building blocks of a distinct species (or set of species), the amount of information required to describe this structure can be taken to be the same as that required to describe an infinite chain consisting of the same elements. A simple example is given in Figure \ref{steric}. The crucial assumption which has to hold for this simplification to work is that the geometry of the loop is specified by the species (and, by extension, the geometry) of the building block. For proteins as building blocks of protein complexes, this is a very reasonable assumption. In the case of molecules it would furthermore be possible to simplify the self-assembly kit by introducing building blocks representing common small loop structures, such as carbon rings. 

{\em Multiple nuclei} -- In principle one could consider beginning the self-assembly with multiple nuclei in place. Multiple nuclei may, through steric hindrance or modular repetition, be used to achieve certain structures in a more efficient way, using fewer building blocks than a single nucleus would require. This reduction in complexity may however be countered in practical applications by the difficulty of achieving the required precise relative displacements of nucleus particles. It is because of these reasons that we have concentrated on a single nucleus model, as the positioning of multiple nuclei makes it much more difficult to construct a general measure of complexity.

\begin{figure}
\begin{center}
\includegraphics[width=6cm]{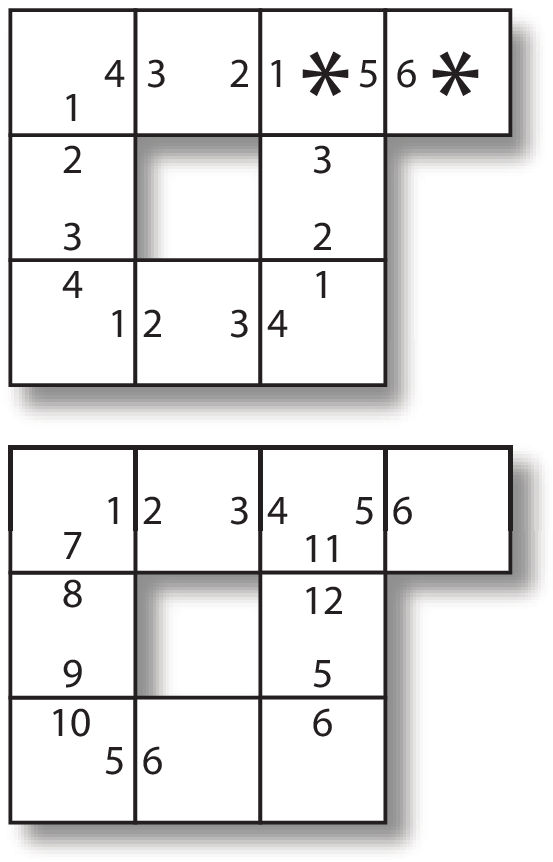}
\caption{ \footnotesize Illustration of nuclei placement. (Top:) If we specify either of the two starred blocks as nuclei, deterministic bonding will result. However, if any other block is used as the nucleus, bonding will be non-deterministic, as both the $\{1, 0, 0, 4\}$ and $\{1, 0, 5, 0\}$ blocks can join the open $2$ edges that will form. This self-assembly kit has a complexity of $K = 42.4$ bits. (Bottom:) A general nucleus system to produce the same structure, illustrating the required increase in complexity ($K = 98.1$ bits).}\label{nuclei}
\end{center}
\end{figure}

Within the single nucleus category, we further distinguish between structure with a \emph{specified nucleus block} and those with \emph{general nucleus blocks}. The former case encompasses those assembly kits which are guaranteed to produce a given output structure if and only if a specified block is used as the nucleus (in other words, this block is placed on the substrate before other blocks are introduced to the system). General-nucleus assembly kits by contrast will form the same output structure regardless of which block is placed first. See Figure \ref{nuclei} for an illustration how specifying a nucleus can reduce the complexity of a assembly kit. 

Which of these classes to employ in a study depends on the motivating context of the self-assembling system under consideration. If modelling assembly in a diffusion-dominated environment, for example, the order in which interacting particles meet cannot be specified, so the general-nucleus model is more appropriate. In a controlled environment where a nucleus can be placed to initiate assembly, the single-nucleus model is applicable. The two cases correspond to different `languages' being used to measure complexity, and so care must be taken in comparative studies to only compare numerical complexity values from within one class.

{\em Kolmogorov complexity} -- Our approach to measuring physical complexity is motivated by the concept of Kolmogorov complexity. It is however important to note that while Kolmogorov complexity itself is uncomputable due to the Halting problem \cite{CoverThomas}, our minimum is not. This is because the runtime of a finite computer program with finite output can be infinite, while the assembly time of a finite shape is always finite \cite{Winfree00}. It is possible to define the actual Kolmogorov complexity of a shape \cite{Winfree06}, but this is uncomputable. Our computable complexity measure $K(A)$ forms a bound on this unattainable quantity, and is dependent on the way in which we encode the description of the assembly kit. It therefore is useful for the analysis, classification and comparison of physical structures, as long as we use a consistent encoding.

\section{Conclusion}

We present a general approach for measuring the physical complexity of any connected structure, using the language of self-assembly. This approach is capable of detecting symmetry and modularity in a given structure, because these features significantly decrease the size of the required self-assembly instruction set. It therefore provides a powerful tool for automated classification and categorization of physical structures. In addition, the connection between self-assembly and complexity is an argument for the ubiquity of modular and symmetric features in biological systems: Since many such systems self-assemble, evolving sets of self-assembly instructions are likely to yield symmetric and modular structures, as the instructions for these are more efficient to evolve.

\end{document}